\documentclass[twocolumn]{aastex62}
\usepackage{amsmath}
\usepackage{booktabs}
\usepackage{fontawesome}
\usepackage{bm}
\usepackage{amsbsy}
\usepackage{fontawesome}

\usepackage{soul, CJK}
\usepackage[utf8]{inputenc}

\usepackage{xspace}
\DeclareFixedFont{\ttb}{T1}{txtt}{bx}{n}{9} 
\DeclareFixedFont{\ttm}{T1}{txtt}{m}{n}{9}  
\usepackage{color}
\definecolor{deepblue}{rgb}{0,0,0.5}
\definecolor{deepred}{rgb}{0.6,0,0}
\definecolor{deepgreen}{rgb}{0,0.5,0}
\usepackage{listings}
\usepackage{gensymb}

\newcommand\pythonstyle{\lstset{
language=Python,
basicstyle=\ttm,
otherkeywords={self},             
keywordstyle=\ttb\color{deepblue},
emph={MyClass,__init__},          
emphstyle=\ttb\color{deepred},    
stringstyle=\color{deepgreen},
frame=tb,                         
showstringspaces=false            %
}}
\lstnewenvironment{pyt}[1][]
{
\pythonstyle
\lstset{#1}
}
{}

\newcommand{\vect}[1]{\mathbf{#1}}

\newcommand{\br}[1]{\mathopen{}\left(#1\right)\mathclose{}}

\newcommand{\abs}[1]{\left|#1\right|}

\newcommand{\deriv}[2]{\frac{\partial{#1}}{\partial{#2}}}

\usepackage{listings}

\graphicspath{{./}{figures/}}

\shorttitle{Real-Time Likelihood-Free Inference of \textit{Roman} Binary Microlensing Events}
\shortauthors{Zhang et al.}

\begin{document}

\title{Real-Time Likelihood-free Inference of \textit{Roman} Binary Microlensing Events with Amortized Neural Posterior Estimation}

\correspondingauthor{Keming Zhang}
\email{kemingz@berkeley.edu}

\author[0000-0002-9870-5695]{Keming Zhang \begin{CJK*}{UTF8}{gkai}(张可名)\end{CJK*}}
\affil{Department of Astronomy, University of California, Berkeley, CA 94720-3411, USA}
\author[0000-0002-7777-216X]{Joshua S. Bloom}
\affil{Department of Astronomy, University of California, Berkeley, CA 94720-3411, USA}
\affil{Lawrence Berkeley National Laboratory, 1 Cyclotron Road, MS 50B-4206, Berkeley, CA 94720-3411, USA}
\author[0000-0003-0395-9869]{B. Scott Gaudi}
\affil{Department of Astronomy, The Ohio State University, Columbus, OH 43210, USA}
\author[0000-0001-7956-0542]{Fran\c{c}ois Lanusse}
\affil{AIM, CEA, CNRS, Universit\'e Paris-Saclay, Universit\'e Paris Diderot, Sorbonne Paris Cit\'e, F-91191, Gif-sur-Yvette, France}
\author[0000-0002-6406-1924]{Casey Lam}
\affil{Department of Astronomy, University of California, Berkeley, CA 94720-3411, USA}
\author[0000-0001-9611-0009]{Jessica R. Lu}
\affil{Department of Astronomy, University of California, Berkeley, CA 94720-3411, USA}

\begin{abstract}
Fast and automated inference of binary-lens, single-source (2L1S) microlensing events with sampling-based Bayesian algorithms (e.g., Markov Chain Monte Carlo; MCMC) is challenged on two fronts: high computational cost of likelihood evaluations with microlensing simulation codes, and a pathological parameter space where the negative-log-likelihood surface can contain a multitude of local minima that are narrow and deep. Analysis of 2L1S events usually involves grid searches over some parameters to locate approximate solutions as a prerequisite to posterior sampling, an expensive process that often requires human-in-the-loop domain expertise. As the next-generation, space-based microlensing survey with the \textit{Roman Space Telescope} is expected to yield thousands of binary microlensing events, a new fast and automated method is desirable.
Here, we present a likelihood-free inference (LFI) approach named amortized neural posterior estimation, where a neural density estimator (NDE) learns a surrogate posterior $\hat{p}(\pmb{\theta}|\vect{x})$ as an observation-parametrized conditional probability distribution, from pre-computed simulations over the full prior space.
Trained on 291,012 simulated \textit{Roman}-like 2L1S simulations, the NDE produces accurate and precise posteriors within seconds for any observation within the prior support without requiring a domain expert in the loop, thus allowing for real-time and automated inference.
We show that the NDE also captures expected posterior degeneracies. The NDE posterior could then be refined into the exact posterior with a downstream MCMC sampler with minimal burn-in steps.

\end{abstract}


\section{Introduction}

When the apparent trajectory of a foreground \textit{lens} star passes close to a more distant \textit{source} star, the gravitational field of the \textit{lens} will perturb the light rays from the \textit{source} which results in a time-variable magnification. Such are single-lens, single-source (1L1S) microlensing events. Binary microlensing events occur when the \textit{lens} is a system of two masses: either a binary star system or a star-planet configuration. Observation of such events provides a unique opportunity for exoplanet discovery as the planet-to-star mass ratio may be inferred from the light curve without having to detect light from the star-planet \textit{lens} itself (see \citealt{gaudi_exoplanetary_2010} for a review). A next-generation microlensing survey with the Roman Space Telescope (\citealt{spergel_wide-field_2015}; hereafter \textit{Roman}) is estimated to discover thousands of binary microlensing events over the duration of the 5-year mission span, many with planetary-mass companions \citep{penny_predictions_2019}, which is roughly an order of magnitude more than events previously discovered (see \citealt{gaudi_microlensing_2012} for a review).

While single-lens microlensing events are described by a simple analytic expression (``Paczy\'nski light-curve''), binary microlensing events require numerical forward models that are computationally expensive. In addition, binary microlensing light-curves exhibit extraordinary phenomenological diversity, owing to the different geometrical configurations for which magnification could take place. This translates to a parameter space for which the likelihood surface suffers from a multitude of local minima that are disconnected, narrow, and deep; this issue significantly hampers any attempt of direct sampling-based inference such as MCMC where the chains are initialized from a broad prior. As a result, binary microlensing events thus far have generally been analyzed on a case-by-case basis.

For some planetary-mass-ratio events, heuristics could be used to ``read off'' an approximate solution from the planetary anomaly in the light curve \citep{gaudi_planet_1997,gould_discovering_1992}. \cite{khakpash_fast_2019} applied the heuristics described in \cite{gaudi_planet_1997} on simulated \textit{Roman} light-curves and found that the projected binary separation can be recovered very well for low-mass-ratio events, and the binary mass-ratios within an order of magnitude for events with wide and close caustic topologies. 

More generally, an expensive grid search is usually conducted over a subset of parameters to which the magnification pattern is hyper-sensitive:  i.e., binary separation, mass ratio, and the source trajectory angle of approach (e.g. \cite{herrera-martin_ogle-2018-blg-0677lb_2020}). At each grid-point, the remaining parameters are searched for with simple Nelder-Mead optimization \citep{nelder_simplex_1965} or MCMC. The fixed-grid solutions are then used to seed full MCMC samplings to refine solutions and sample the posteriors. This status quo approach, which is both computationally expensive and requires domain expertise in the loop, thus presents a great challenge to analyze the thousands of binary microlensing events expected to be discovered by \textit{Roman}.

Recent progress in deep learning provides a promising path for a solution. In particular, both Convolutional (CNN; \citealt{lecun_deep_2015}) and Recurrent Neural Networks (RNN \citealt{hochreiter_long_1997,cho_learning_2014}) have emerged as powerful alternatives to feature engineering of astronomical time-series (e.g. \citealt{naul_recurrent_2018}). Given sufficient training data, CNN/RNNs could learn to compress the ``high-dimensional'' raw observations into ``low-dimensional'' feature vectors---automatically learning to produce features that are useful for downstream tasks such as classification or regression. \cite{vermaak_rapid_2003} applied a more basic form of the neural network --- the multilayer perceptron (MLP) --- to predict for 2L1S parameters on simulated noise-free light-curves, and achieved a success rate of 68\% when the MLP results were further refined with Nelder-Mead optimization \citep{nelder_simplex_1965}. However, there remains a large gap between the proof-of-concept work of \cite{vermaak_rapid_2003} and application to real data due to the omission of noise and restrictions in parameter space.
Additionally, machine learning has also been previously applied to \textit{discover} and \textit{classify} microlensing events \citep{wyrzykowski_ogle-iii_2015,godines_machine_2019,mroz_identifying_2020}.

In addition to advances in this ``representation learning,'' neural networks have also enjoyed significant progress in modeling probability distributions, otherwise known as neural density estimation, where the fundamental task is learn distributions from samples of that distribution. Both autoregressive models \citep{germain_made_2015,oord_wavenet_2016} and flow-based models \citep{papamakarios_masked_2017,dinh_density_2017} are NDEs that are highly capable of modeling complicated and multi-modal distributions, which can not only evaluate probability densities, but also sample from that distribution. NDEs thus allow for flexible uncertainty quantification and degenerate solutions which were not possible in \cite{vermaak_rapid_2003}.

The advancement in feature learning and NDE has allowed for accelerated progress in the field of likelihood-free inference (LFI), also known as simulation-based inference (SBI), which has been motivated by inference problems without a tractable likelihood. LFI is an umbrella term that encompasses a wide range of inference algorithms that do not require explicit evaluation of the likelihood. Under our particular LFI approach called amortized neural posterior estimation, an NDE learns a surrogate posterior as an observation-parametrized conditional probability distribution, from pre-computed simulations over the full prior space. A ``featurizer'' neural network is employed to compress raw observation into a feature vector which parametrizes the NDE.
Inference is amortized in that all of the computation cost of simulation is paid upfront---likelihood evaluation with the slow forward simulator is no longer required, thus allowing for fast inference.
For other neural LFI instances, neural networks could learn the likelihood \citep{papamakarios_sequential_2019} or the likelihood-ratio \citep{thomas_likelihood-free_2020} as surrogates to accelerate sampling-based inference algorithms like MCMC (see \citealt{cranmer_frontier_2020} for an overview).

In this paper, we present a likelihood-free inference approach for binary microlensing where an NDE learns a surrogate posterior $\hat{p}(\pmb{\theta}|\vect{x})$ as an observation-parametrized conditional distribution from $(\vect{x}^{i}, \pmb{\theta}^{i})$ samples of simulated microlensing light-curves with the associated microlensing parameters. After training, the NDE can automatically generate posterior samples for future observations effectively in real-time. Because of the speed and performance without supervision by domain experts, the approach we introduce here has great potential for batch inference tasks such as those posed by \textit{Roman}. 
Our preliminary results were reported as an extended abstract in \cite{zhang_automating_2020}. The work herein supersedes and expands upon that work. 

We first lay out our inference framework in Section \ref{sec:infer}. Training set construction under the context of \textit{Roman} is discussed in Section \ref{sec:data}. In Section \ref{sec:results}, we demonstrate the ability of the NDE to capture degenerate solutions and also present a systematic evaluation of the NDE performance over a large number of test events. In Section \ref{sec:discussion}, we suggest future directions including a potential addition of a down-stream MCMC algorithm to refine the NDE posterior into the exact posterior, with minimal additional computation time.

\section{Method}
\label{sec:infer}

\begin{figure*}
    \includegraphics[width=\textwidth]{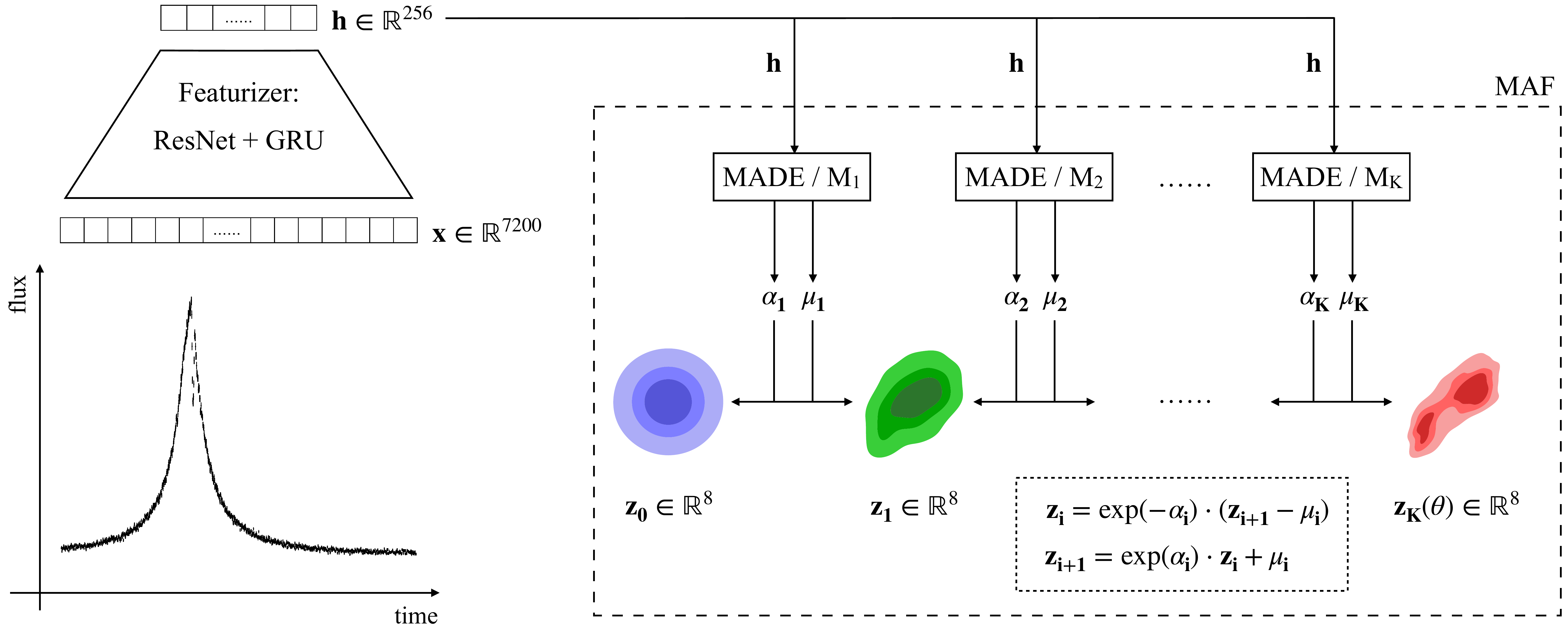}
    \caption{Schematic illustration of the inference framework based on conditional NDE. The bottom left shows a microlensing light-curve in arbitrary units which is abstracted into the length-7200 vector ($\vect{x}$) above. The featurizer composed of a combination of ResNet and GRU, shown in the trapezoid, compresses the light-curve into a low-dimensional feature vector $\vect{h}$. The masked autoregressive flow (MAF), composed of $K$ blocks of masked autoencoder for density estimation (MADE), is shown in the dashed box. Each MADE block takes in the feature vector $\vect{h}$ and predicts scaling ($\pmb{\alpha}$) and shifting ($\pmb{\mu}$) factors, which parameterizes an invertible affine transformation between adjacent random variables (e.g., $\vect{z}_0$ and $\vect{z}_1$) shown in the dotted box. The left-most random variable is the mixture-of-Gaussian base distribution whereas the right-most random variable ($\vect{z}_K$) is the posterior ($\pmb{\theta}$).}
    \label{fig:network}
\end{figure*}

NDEs are neural networks that are capable of learning distributions from samples.
We train an NDE to learn a surrogate posterior $\hat{p}(\pmb{\theta}|\vect{x})$ as an observation-parametrized conditional distribution from $(\vect{x}^{i}, \pmb{\theta}^{i})$ samples of simulated microlensing light-curves, where $\pmb{\theta}^i$ are the physical parameters and $\vect{x}^{i}\in \mathbb{R}^{\rm N}$ is the light curve with $N$ data-points. The training objective is to minimize the Kullback--Leibler (KL) divergence ($D_{\rm KL}$), or relative entropy, which is a measure of how one probability distribution ($Q$) is different from a reference probability distribution ($P$):
\begin{equation}
    D_{\rm KL}(P||Q) = \mathbb{E}_{x\sim p(x)}\left[\log\left(\frac{p(x)}{q(x)}\right)\right]
\end{equation}
In this case, we would like to minimize the KL divergence from the NDE surrogate posterior $\hat{p}(\pmb{\theta}|\vect{x})$ to the true posterior $p(\pmb{\theta}|\vect{x})$:
\begin{flalign}
    \phi&={\rm argmin}(D_{\rm KL}(p(\pmb{\theta}|\vect{x})) || \hat{p}_{\phi}(\pmb{\theta}|\vect{x})))\nonumber\\
    &={\rm argmin}(\mathbb{E}_{\pmb{\theta}\sim p(\pmb{\theta}), \vect{x}\sim p(\vect{x}|\pmb{\theta})}[\log(p(\pmb{\theta}|\vect{x}))-\log(\hat{p}_{\phi}(\pmb{\theta}|\vect{x}))])\nonumber\\
    &={\rm argmax}(\mathbb{E}_{\pmb{\theta}\sim p(\pmb{\theta}), \vect{x}\sim p(\vect{x}|\pmb{\theta})}[\hat{p}_{\phi}(\pmb{\theta}|\vect{x})]),
    \label{eq:kl}
\end{flalign}
where $\phi$ represents the neural network parameter, and $\mathbb{E}$ denotes the mathematical expectation over the specified distribution.

In light of Equation \ref{eq:kl}, the NDE is therefore trained through Maximum Likelihood Estimation (MLE)
on a training set with physical parameters drawn from the prior $p(\pmb{\theta})$ and light-curves drawn from the likelihood function, which is the Poisson measurement noise model on top of the noise-free microlensing light curve $g(\pmb{\theta})$ (in the number of photons) which, for simplicity, is assumed to be in the Gaussian limit:
\begin{equation}
    p(\vect{x}|\pmb{\theta})=\mathcal{N}\left(\mu=g(\pmb{\theta}), \sigma=\sqrt{g(\pmb{\theta})}\right).
\label{eq:noise}
\end{equation}

The noise-free light-curve, in turn, is determined by the baseline \textit{source} flux ($F_{\rm source}$), the magnification time-series produced by the microlensing physical forward model $A(\pmb{\theta})$, and the constant \textit{blend} flux, which is the flux from the \textit{lens} star and any other star that is unresolved from the source star:
\begin{equation}
    g(\pmb{\theta})=A(\pmb{\theta})\cdot F_{\rm source}+F_{\rm blend}.
\end{equation}

We use a 20-block Masked Autoregressive Flow (MAF) \citep{papamakarios_masked_2017} to model $\hat{p}(\pmb{\theta}|\vect{x})$, and a ResNet-GRU network to extract features ($\vect{h}$) from the light curve ($\vect{x}$). We do not distinguish between $\hat{p}(\pmb{\theta}|\vect{x})$ and $\hat{p}(\pmb{\theta}|\vect{h})$ where the former is meant to refer to the ``featurizer+NDE'' model and the latter is meant to refer to the NDE model alone that is explicitly conditioned on $\vect{h}$. Figure \ref{fig:network} presents a diagram of our neural posterior estimation framework. The ResNet-GRU network is comprised of a 18-layer 1D ResNet (Residual Convolutional Network; \citealt{he_deep_2016}) and a 2-layer GRU (Gated Recurrent Network; \citealt{cho_learning_2014}). We describe the neural networks in detail below.

\subsection{Masked Autoregressive Flow}
The masked autoregressive flow (MAF) belongs to a class of NDE called normalizing flows, which models the conditional distribution $\hat{p}(\pmb{\theta}|\vect{x})$ as an invertible transformation $f$ from a base distribution $\pi_z\br{\vect{z}}$ to the target distribution $\hat{p}(\pmb{\theta}|\vect{x})$. The base density $\pi_u\br{\vect{z}}$ is required to be fast to evaluate and is typically chosen to be either a standard Gaussian or a mixture of Gaussians for the MAF. The basic idea is that if the MAF, conditioned on the observation $\vect{x}$, could learn to map the posterior to a standard Gaussian, then the inverse transformation could enable sampling of the posterior by simply sampling from that standard Gaussian.

As binary microlensing events often exhibit degenerate, multi-modal solutions, we use a mixture of eight standard multivariate Gaussians, each with 8 dimensions, as the base distribution. The posterior probability density $\hat{p}(\pmb{\theta}|\vect{x})$ is evaluated by applying the inverse transformation $f^{-1}$ from $\pmb{\theta}$ to $\vect{z}$:
\begin{equation}
\hat{p}(\pmb{\theta}|\vect{x}) = \pi_z\br{f^{-1}\br{\pmb{\theta}}}\abs{\det\br{\deriv{f^{-1}}{\pmb{\theta}}}},
\end{equation}
where $\pi_z\br{f^{-1}\br{\pmb{\theta}}}$ represents the probability density for the base distribution ($\pi_z$) evaluated at $f^{-1}\br{\pmb{\theta}}$, while the second term---the determinant of the Jacobian---corresponds to the ``compression'' of probability space. 

The MAF is built upon blocks of affine transformations where the scaling and shifting factors for each dimension are computed with a Masked Autoencoder for Distribution Estimation (MADE; \citealt{germain_made_2015}). For a simple 1-block case, the inverse transformation from $\pmb{\theta}$ to $\vect{z}$ is expressed as:

\begin{equation}
\label{eq:infer}
z_i = (\theta_i - \mu_i)\cdot\exp{(-\alpha_i)},
\end{equation}
In the above equation, 
\begin{align}
    \mu_i = f_{\mu_i}\br{\pmb{\theta}_{1:i-1}; \vect{x}}\\
    \alpha_i = f_{\alpha_i}\br{\pmb{\theta}_{1:i-1}; \vect{x}}
\end{align}
are the scaling and shifting factors modeled by MADE subject to the autoregressive condition that the transformation of any dimension can only depend on those prior to it according to a predetermined ordering.
This allows the Jacobian of $f^{-1}$ to be triangular, whose absolute determinant can be easily calculated as:
\begin{equation}
\label{eq:abs_det}
\abs{\det\br{\deriv{f^{-1}}{\pmb{\theta}}}} = \exp\br{-{\sum}_i\alpha_i},
\end{equation}
where $\alpha_i = f_{\alpha_i}\br{\pmb{\theta}_{1:i-1}; \vect{x}}$. 

To sample from the posterior, the forward transformation $\pmb{\theta}=f(\vect{z})$ where $\vect{z}\sim\pi_{z}$ is applied:
\begin{equation}
\theta_i = z_i\cdot\exp{\alpha_i} + \mu_i,
\end{equation}
where $\mu_i$ and $\alpha_i$ are computed in the same manner as the inverse transformation.

The MAF is built by stacking many such affine transformation blocks, $M_1, M_2, \ldots, M_K$, where $M_K$ models the invertible transformation $f_K$ between the posterior ($\vect{z}_K$) and intermediate random variable $\vect{z}_{K-1}$, $M_{K-1}$ models that between intermediate random variables $\vect{z}_{K-1}$ and $\vect{z}_{K-2}$ and so on, and finally the base random variable $\vect{z}_0$ is modeled with the mixture-of-Gaussian distribution. $M_1$ also computes the mixture weights. The composite transformation can be written as $f=f_{1}\circ f_{2}\circ\ldots\circ f_{K}$ and the posterior probability density is now:
\begin{flalign}
\hat{p}(\pmb{\theta}|\vect{x}) &= \pi_z\br{f^{-1}\br{\pmb{\theta}}}\prod_{i=1}^{K}\abs{\det\br{\deriv{f_{i}^{-1}}{\vect{z}_{i-1}}}}
\end{flalign}
where it is understood that $\vect{z}_{K}:=\pmb{\theta}$. The log-probability of the posterior is then, by Equation \ref{eq:abs_det},

\begin{flalign}
\log\hat{p}(\pmb{\theta}|\vect{x}) 
&= \log[\pi_z\br{f^{-1}\br{\pmb{\theta}}}] + \sum_{i=1}^{K}\log\abs{\det\br{\deriv{f_{i}^{-1}}{z_{i-1}}}}\nonumber\\
&= \log[\pi_z\br{f^{-1}\br{\pmb{\theta}}}] - \sum_{i=1}^{K}\sum_{j=1}^{N}\alpha_{j}^{i}\nonumber,\\
\end{flalign}
where $\alpha_{j}^{i}$ is the $j$th component of the scale factor in $M_{i}$, as in Equation \ref{eq:infer}. This serves as the optimization objective (see Section \ref{sec:train}).
 
Autoregressive models are sensitive to the order of the variables. The original MAF paper uses the default order for the autoregressive layer closest to $\pmb{\theta}$ and reverses the order for each successive layer. In this work, we adopt fixed random orderings for each MAF block which we find to allow for better expressibility. The random seed of the ordering serves as a hyper-parameter to be optimized on.

\subsection{Featurizer Network}

A custom 1D ResNet with a down-stream 2-layer GRU is used as the light curve featurizer which takes preprocessed light curves ($\vect{x}$) as input and outputs a low-dimensional feature vector ($\vect{h}$). The ResNet used in this study shares the identical architecture as \cite{zhang_classification_2020} (except for hyper-parameters) and consists of 9 identical residual blocks, each of which is composed of two convolutions followed by layer normalization \citep{ba_layer_2016}. A residual connection is added between each adjacent residual block, which acts as a ``gradient highway'' to assist network optimization. A \texttt{MaxPool} layer is applied in between every two ResNet layers, where the sequence length is reduced by half and the feature dimension doubled until a specified maximum. This results in an output feature map of length $L=56$ and dimension $D=256$, when is then fed into the GRU network that sequentially processes information across the temporal dimension and outputs a single vector of $D=256$ which then serves as the conditional input to the MAF.

\section{Data}
\label{sec:data}
Training data is generated within the context of the Roman Space Telescope Cycle-7 design (see \citealt{penny_predictions_2019}). We first simulate 10$^{6}$ 2L1S magnification sequences with the microlensing code \texttt{MulensModel} \citep{poleski_modeling_2019}; each sequence contains 144 days at a cadence of 0.01 day, corresponding to the planned Roman cadence of 15 minutes \citep{penny_predictions_2019}. These sequences are chosen to have twice the length of the 72-day Roman observation window to facilitate sampling from a $t_0\sim \rm Uniform(0,72)$ prior (see Section \ref{sec:prior}). We then fit each simulated magnification time-series with a Paczy\'nski single-lens-single-source (1L1S) model (assuming $S/N_{\rm base}=200$ and $f_s=1$; see Section \ref{sec:lc}) and discard those that are consistent with 1L1S ($\chi^2/\rm dof<1$). This results in a final dataset of 291,012 light curves, among which 95\% (276,461) are used as training set and the remaining 5\% (14,551) as test set.

\subsection{Prior}
\label{sec:prior}
Assuming rectilinear relative motion of the observer, lens, and source, binary microlensing (2L1S) events are characterised by eight parameters: binary lens separation ($s$), mass ratio ($q$), angle of the source trajectory with respect to the projected binary lens axis ($\alpha$), impact parameter ($u_0$), time of closest approach ($t_0$), Einstein ring crossing timescale ($t_E$), finite source size ($\rho$), and source flux fraction ($f_s$). $\alpha$ is the angle between the vector pointing from the primary to the secondary and the source trajectory vector, measured counterclockwise in degrees. $u_0$ and $t_0$ are defined with respect to the binary lens center-of-mass (COM). Where applicable, the parameters are normalized to the Einstein ring length-scale or the Einstein ring crossing time-scale of the total mass of the lens system. $t_0$ and $t_E$ are in units of days. We simulate 2L1S events based on the following analytic priors:
\begin{gather}
    \label{eq:prior}
    s \sim {\rm LogUniform}(0.2, 5)\nonumber\\
    q \sim {\rm LogUniform}(10^{-6}, 1)\nonumber\\
    \alpha \sim {\rm Uniform}(0, 360)\\
    u_0 \sim {\rm Uniform}(0, 2)\nonumber\\
    t_0 \sim {\rm Uniform}(0, 72)\nonumber\\
    t_E \sim {\rm TruncLogNorm}(1, 100, \mu=10^{1.15}, \sigma=10^{0.45})\nonumber\\
    \rho \sim {\rm LogUniform}(10^{-4}, 10^{-2})\nonumber\\
    f_s\sim \rm LogUniform(0.1,1)\nonumber
\end{gather}

\begin{figure}
    \includegraphics[width=\linewidth]{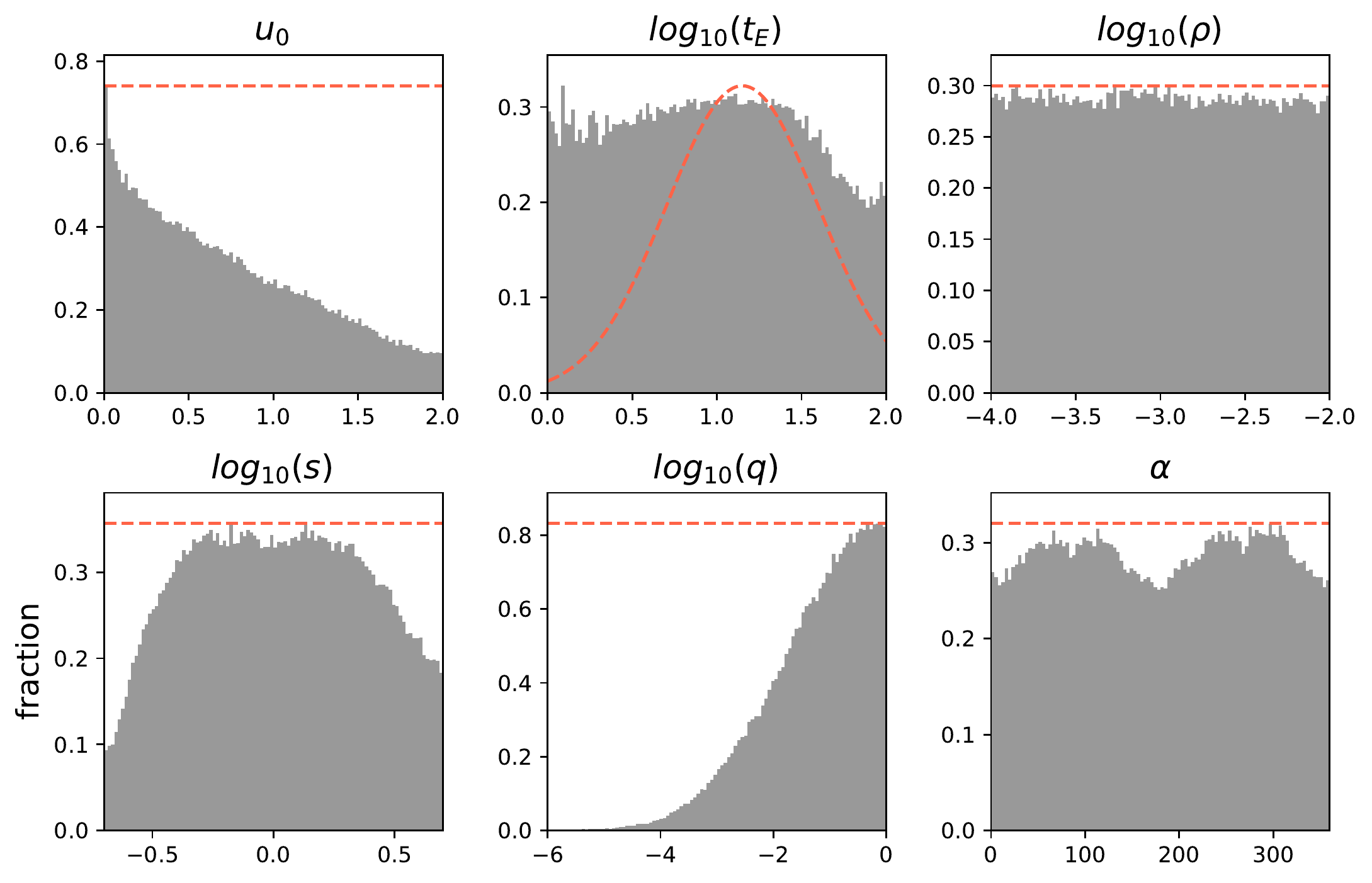}
    \caption{Fraction of the $10^6$ simulations passing the $\chi^2_{1L1S}/\rm dof>1$ cutoff as a function of each parameter, shown in the gray histograms. The original analytic priors used to generate the $10^6$ simulations are shown in red-dashed lines up to a normalization factor. For parameters with a flat original prior, the gray histogram is also the effective training set prior up to a normalization factor. The $t_0$ and $f_s$ distributions follow the original priors as they are sampled on the fly during training.}
    \label{fig:param}
\end{figure}

We note that because of the $\chi^2_{1L1S}/\rm dof<1$ cutoff, the effective prior is the parameter distribution for the 276,461 training set simulations, different from the prior above. As shown in Figure \ref{fig:param}, large $\log{q}$ and small $u_0$, which otherwise have flat priors, are strongly preferred.

During training, a random 72-day segment is chosen on the fly from each 144-day magnification sequence, equivalent to prescribing a uniform prior on $t_0$.
The truncated normal distribution for $t_E$ is an approximation of a statistical analysis based on OGLE-IV data \citep{mroz_no_2017}. The lower limit of $q = 10^{-6}$ corresponds to the mass ratio between Mercury and a low-mass ($M\sim 0.1 M_{\odot}$) M-dwarf star, highlighting the superb sensitivity of Roman. 
The source flux fraction is defined as the ratio between the \textit{source} flux and the total baseline flux
\begin{equation}
    f_s=\dfrac{F_{\rm source}}{F_{\rm source}+{F_{\rm blend}}}.
\end{equation}

\subsection{Light-curve realization}
\label{sec:lc}
The magnification sequences are converted into light-curves during training on the fly by multiplying with the baseline pre-magnification \textit{source} flux before adding the constant \textit{blend} flux and applying measurement noise. 
For simplicity, we only consider photon-counting noise from the lens and fixed blend flux, assumed to be in the Gaussian limit of the Poisson noise (Equation \ref{eq:noise}), where the standard deviation of each photometric measurement is the square root of flux measurement in photon counts. Studies of the bulge star population show that the apparent magnitude largely lies within the range of 20 mag to 25 mag (\citealt{penny_predictions_2019}: Figure 5). The Roman/WFIRST Cycle 7 design has the zero-point magnitude (1 count/s) at 27.615 mag for the W149 filter. With exposure time at 46.8\,s, the aforementioned magnitude range corresponds to signal-to-noise ($S/N_{\rm base}$) ratios between 230 and 23 for the baseline flux, which we randomly and uniformly sample during training. On-the-fly sampling of $S/N_{\rm base}$ and $f_s$ also serves as data augmentation, which refers to the process of expanding the effective size of the training set.

\subsection{Pre-processing and Training}
\label{sec:train}

Network optimization is performed with ADAM \citep{kingma_adam:_2015} at an initial learning rate of 0.001 and batch size 512, which decays to 0 according to a cosine annealing schedule \citep{loshchilov_sgdr_2017} for 250 epochs, at which point the training terminates. To ensure that there is no over-fitting, we first reserved 20\% of the training set as a validation set. After confirming the absence of over-fitting, we then proceed with the full training set. We apply data augmentation on $\alpha$ by changing the direction of the source trajectory: the temporal order of each sequence is reverted and $\alpha$ becomes $-(\alpha + 180) \mod 360$. Each training epoch takes $\sim6$ minutes on four NVidia GTX 2080 Ti GPUs with a total training time of around 25 hours. As an evaluation metric, the final average negative log-likelihood (NLL) is $-$16.316 on the training set and $-$16.177 on the test set, where a lower value represents a better model fit to the data.

\begin{figure*}
 \begin{center}
    \includegraphics[width=0.95\textwidth]{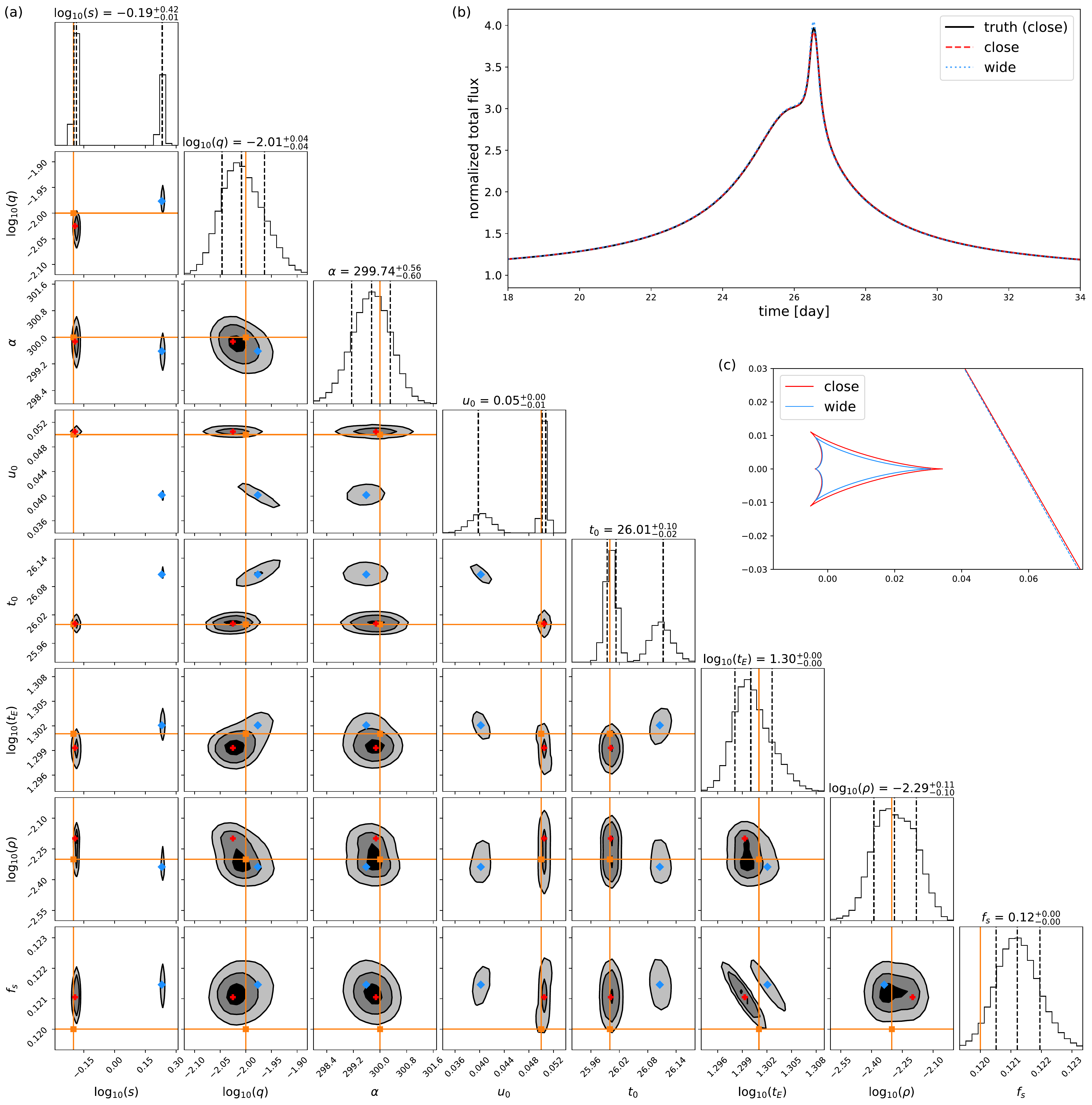}
    \caption{(a) NDE posterior for a central-caustic passing event. $t_E$ and $t_0$ are in units of days, $\alpha$ in degrees, $u_0$, $s$, and $\rho$ in units of $\theta_E$. Filled contours show 1/2/3$\sigma$ regions. The ground truth close solution is marked with orange cross-hairs. The close and wide solutions are marked with a red cross and a blue diamond, respectively. (b) Close-up view of the light-curve realizations normalized to the minimum fluxes for both solutions, in the same color-coding as the left panel. The 0.01 day cadence and measurement noise is negligibly small on the scale of the figure, and therefore not shown. (c) Caustic structures as well as trajectories for the two solutions in the same color-coding, centered on the center of caustic.}
    \label{fig:posterior}
    \end{center}
\end{figure*}

\begin{figure*}
 \begin{center}
    \includegraphics[width=0.9\textwidth]{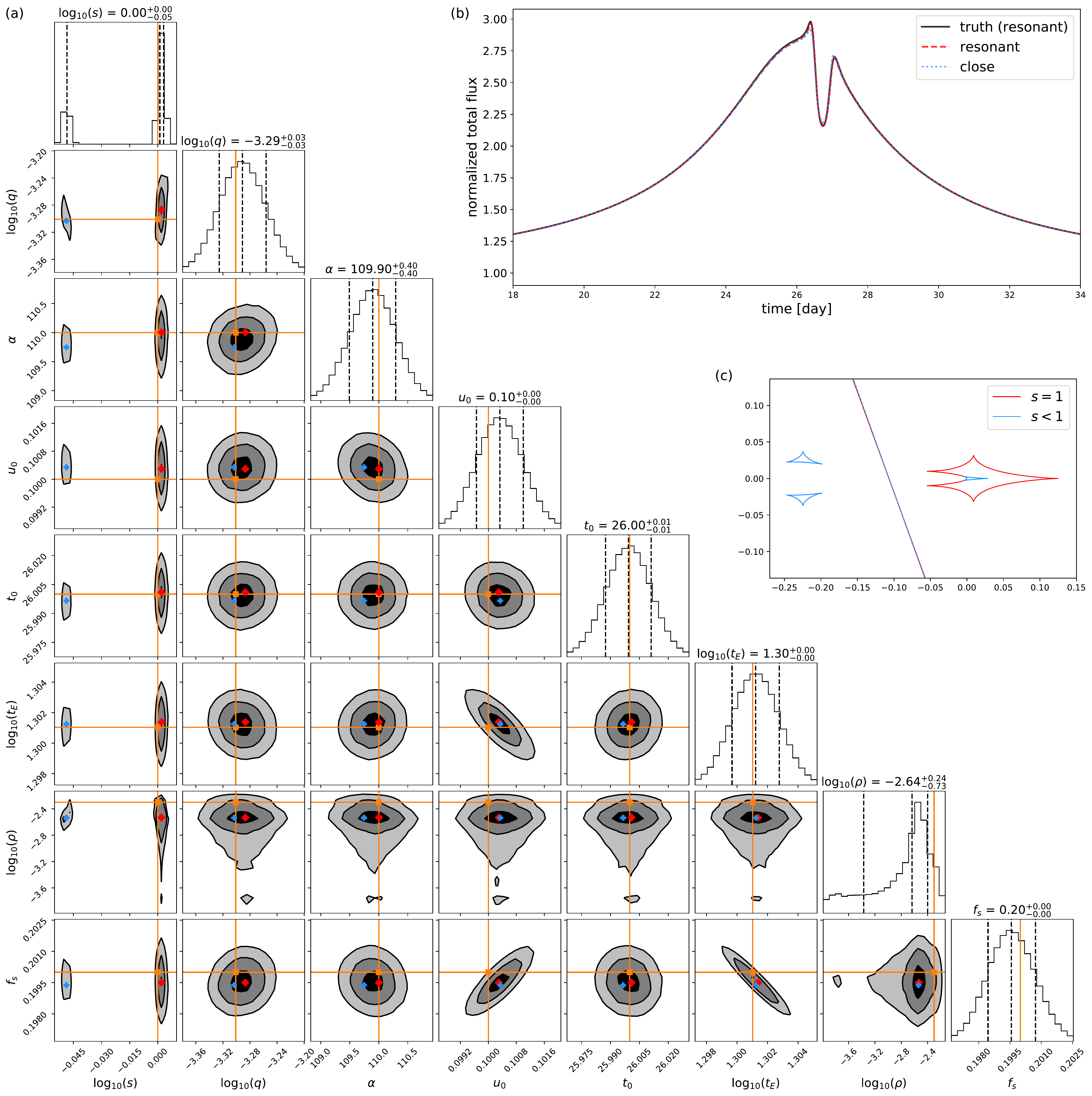}
    \caption{Resonant-caustic-passing event; same figure caption as Figure \ref{fig:posterior}. Here, a degenerate solution is seen at $s<1$, whose two triangular caustics cause a similar suppression pattern as the resonant caustic.}
    \label{fig:posterior-resonant}
    \end{center}
\end{figure*}

\begin{figure*}
 \begin{center}
    \includegraphics[width=\textwidth]{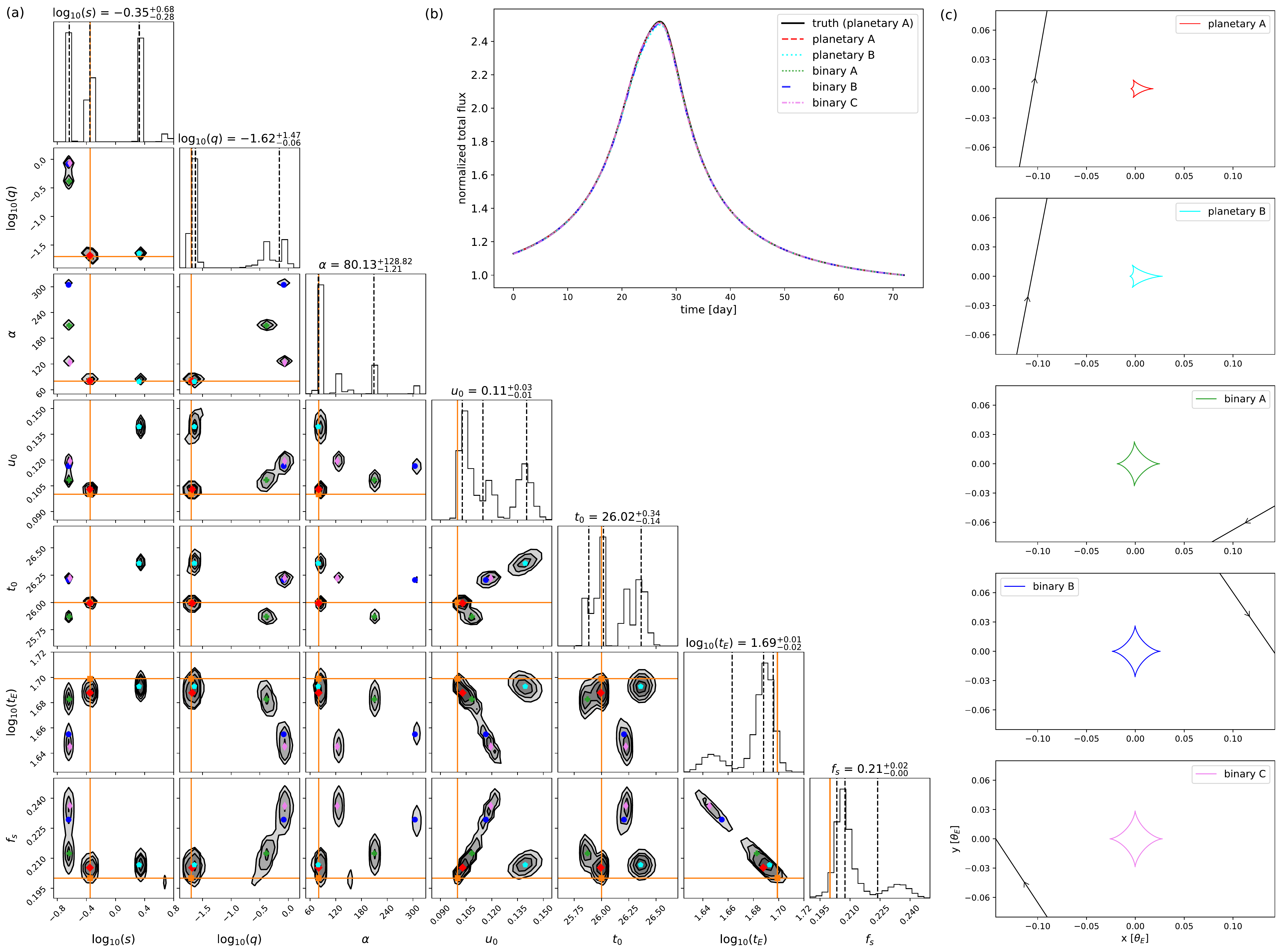}
    \caption{Example event exhibiting a blunt and flat light-curve near the peak, which has a 5-fold degenerate NDE posterior; same figure caption as Figure \ref{fig:posterior} for (a) and (b). (c) Caustic structures and source trajectories for the five solutions.} The same color-coding is shared across the three panels.
    \label{fig:posterior-binary}
    \end{center}
\end{figure*}

\begin{deluxetable}{cc|cc}
    \tablecaption{Solutions for the example central-caustic passing event. $t_E$ and $t_0$ are in units of days, $\alpha$ in degrees, $u_0$, $s$, and $\rho$ in units of $\theta_E$. Uncertainties are $1\sigma$ marginal uncertainties.\label{tab:param-central}}
    \tablehead{\colhead{} & \colhead{truth} & \colhead{close} & \colhead{wide}}
    \startdata
    $\log_{10}(s)$ & $-0.200$ &$-0.1923_{-0.0036}^{+0.0034}$ &$0.2301_{-0.0040}^{+0.0049}$ \\
    $\log_{10}(q)$ & $-2.000$ &$-2.0252_{-0.0147}^{+0.0144}$ &$-1.9761_{-0.0155}^{+0.0177}$ \\
    $\alpha$ & $300.000$ &$299.8524_{-0.2658}^{+0.2457}$ &$299.5919_{-0.3139}^{+0.2469}$ \\
    $u_0$ & $0.050$ &$0.0505_{-0.0002}^{+0.0002}$ &$0.0403_{-0.0010}^{+0.0007}$ \\
    $t_0$ & $26.000$ &$26.0027_{-0.0063}^{+0.0051}$ &$26.1054_{-0.0075}^{+0.0131}$ \\
    $\log_{10}(t_E)$ & $1.301$ &$1.2993_{-0.0008}^{+0.0007}$ &$1.3020_{-0.0009}^{+0.0010}$ \\
    $\log_{10}(\rho)$ & $-2.301$ &$-2.2075_{-0.1133}^{+-0.0044}$ &$-2.3421_{-0.0210}^{+0.0737}$ \\
    $f_s$ & $0.120$ &$0.1211_{-0.0003}^{+0.0003}$ &$0.1214_{-0.0003}^{+0.0004}$ \\
    \enddata
\end{deluxetable}

\begin{deluxetable}{cc|cc}
    \tablecaption{Solutions for the example resonant-caustic passing event. Same units as Table \ref{tab:param-central}. Uncertainties are $1\sigma$ marginal uncertainties.\label{tab:param-resonant}}
    \tablehead{\colhead{} & \colhead{truth} &\colhead{close}& \colhead{resonant}}
    \startdata
    $\log_{10}(s)$ & $0.000$ &$-0.0484_{-0.0006}^{+0.0017}$ &$0.0018_{-0.0003}^{+0.0010}$ \\
    $\log_{10}(q)$ & $-3.301$ &$-3.3008_{-0.0201}^{+0.0098}$ &$-3.2858_{-0.0121}^{+0.0194}$ \\
    $\alpha$ & $110.000$ &$109.7839_{-0.1526}^{+0.2044}$ &$109.9666_{-0.2093}^{+0.1657}$ \\
    $u_0$ & $0.100$ &$0.1004_{-0.0003}^{+0.0003}$ &$0.1003_{-0.0003}^{+0.0003}$ \\
    $t_0$ & $26.000$ &$25.9980_{-0.0064}^{+0.0048}$ &$26.0014_{-0.0062}^{+0.0047}$ \\
    $\log_{10}(t_E)$ & $1.301$ &$1.3012_{-0.0008}^{+0.0007}$ &$1.3012_{-0.0007}^{+0.0008}$ \\
    $\log_{10}(\rho)$ & $-2.301$ &$-2.5335_{-0.2505}^{+0.0492}$ &$-2.5325_{-0.4117}^{+-0.0041}$ \\
    $f_s$ & $0.200$ &$0.1996_{-0.0005}^{+0.0006}$ &$0.1995_{-0.0005}^{+0.0006}$ \\
    \enddata
\end{deluxetable}

\begin{deluxetable*}{cc|cccccc}
    \tablewidth{\textwidth}
    \tablecaption{Degenerate solutions for the binary-planetary degenerate event shown in Figure \ref{fig:posterior-binary}. Same units as Table \ref{tab:param-central}. Uncertainties are $1\sigma$ marginal uncertainties. \label{tab:param-binary}}
    \tablehead{\colhead{} & \colhead{truth} & \colhead{planetary A}& \colhead{planetary B}& \colhead{binary A}& \colhead{binary B}& \colhead{binary C}}
    \startdata
    $\log_{10}(s)$ & $-0.350$ &$-0.3520_{-0.0049}^{+0.0037}$ &$0.3242_{-0.0035}^{+0.0035}$ &$-0.6373_{-0.0047}^{+0.0037}$ &$-0.6450_{-0.0030}^{+0.0026}$ &$-0.6267_{-0.0043}^{+0.0046}$ \\
    $\log_{10}(q)$ & $-1.700$ &$-1.6849_{-0.0140}^{+0.0275}$ &$-1.6464_{-0.0110}^{+0.0190}$ &$-0.3729_{-0.0273}^{+0.0250}$ &$-0.0813_{-0.0250}^{+0.0297}$ &$-0.0609_{-0.0244}^{+0.0162}$ \\
    $\alpha$ & $80.000$ &$80.0207_{-0.3531}^{+0.2170}$ &$79.0411_{-0.3430}^{+0.2682}$ &$209.7867_{-0.8053}^{+0.8607}$ &$304.7107_{-0.6557}^{+0.6110}$ &$123.4429_{-1.3218}^{+0.2513}$ \\
    $u_0$ & $0.100$ &$0.1027_{-0.0006}^{+0.0009}$ &$0.1390_{-0.0016}^{+0.0022}$ &$0.1082_{-0.0011}^{+0.0010}$ &$0.1160_{-0.0011}^{+0.0012}$ &$0.1197_{-0.0013}^{+0.0009}$ \\
    $t_0$ & $26.000$ &$25.9926_{-0.0070}^{+0.0084}$ &$26.3604_{-0.0169}^{+0.0245}$ &$25.8701_{-0.0080}^{+0.0089}$ &$26.2091_{-0.0128}^{+0.0065}$ &$26.2230_{-0.0101}^{+0.0102}$ \\
    $\log_{10}(t_E)$ & $1.699$ &$1.6874_{-0.0016}^{+0.0035}$ &$1.6927_{-0.0022}^{+0.0024}$ &$1.6824_{-0.0025}^{+0.0038}$ &$1.6550_{-0.0030}^{+0.0037}$ &$1.6443_{-0.0022}^{+0.0045}$ \\
    $f_s$ & $0.200$ &$0.2048_{-0.0010}^{+0.0018}$ &$0.2065_{-0.0011}^{+0.0015}$ &$0.2114_{-0.0014}^{+0.0027}$ &$0.2280_{-0.0015}^{+0.0030}$ &$0.2357_{-0.0021}^{+0.0023}$ \\
    \enddata
\end{deluxetable*}

\section{Results}
\label{sec:results}
The trained model is able to generate accurate and precise posterior samples at a rate of $10^5$ per second on one GPU, effectively in real-time. This is much faster compared to the $\sim1$ per second simulation speed of the forward model \texttt{MulensModel} on one CPU core. In this section, we first highlight the ability of the NDE to capture multi-modal solutions by providing NDE posteriors of representative events where we set the baseline $S/N_{\rm base}=200$. Then, the quality of NDE posteriors is systematically analyzed by examining the accuracy and calibration properties on a test set of 14,511 simulated light-curves.

\subsection{Central-Caustic Passing Event}
\label{sec:central}
Figure \ref{fig:posterior}a shows the NDE posterior for an example central-caustic-passing event where a classic ``close-wide'' degeneracy is clearly exhibited by the $s$-$1/s$ behavior \citep{griest_use_1998,dominik_binary_1999}. Table \ref{tab:param-central} presents the ground truth 2L1S parameters of this event as well as the ``close'' and ``wide'' solutions, calculated as the modes of their respective distributions. The fact that $f_s$ is slightly underestimated is related to a systematic effect as discussed in Section \ref{sec:scatter}. Although the source is expected to pass the caustic center at the same distances for the two cases, Figure \ref{fig:posterior}a shows a bimodal solution for $u_0$ as well because $u_0$ has been defined with respect to the center-of-mass (COM), rather than the caustic center. While the caustic center is centered on the COM for close-separation events, for wide-separation events there is an offset from the COM of
\begin{equation}
\label{eq:offset}
    \delta = \dfrac{s\cdot q}{1+q} - \dfrac{q}{s\cdot (1+q)}
\end{equation}
where the first term accounts for the offset of the caustic center from the location of the primary \citep{han_distinguishing_2008}, and the second term, the offset of the primary from the center of mass. Positive offsets are directed toward the companion and vice versa. Plugging in the wide solution, we expect an offset of $\Delta u_0=0.0116$, which is close to the actual $\Delta u_0 = 0.0099$. Magnification curves of the two solutions, as well as the ground truth are plotted in Figure \ref{fig:posterior}b, which are hardly distinguishable from one another. Figure \ref{fig:posterior}c shows the caustic structures of the two degenerate solutions.

\subsection{Resonant-Caustic Passing Event}

We also highlight an example of a resonant-caustic passing event, whose parameters and solutions are shown in Table \ref{tab:param-resonant}. As illustrated in Figure \ref{fig:posterior-resonant}, the NDE finds an additional solution at $s<1$, whose triangular caustics are causing a similar weak de-magnification as the resonant caustics (also see Figure 7 in \citealt{gaudi_exoplanetary_2010}). This type of degeneracy has been previously observed in the microlensing event OGLE-2018-BLG-0677Lb \citep{herrera-martin_ogle-2018-blg-0677lb_2020}. 
Additionally, strong covariances are seen among $u_0$, $t_E$, and $f_s$, as is also seen in the previous example (Section \ref{sec:central}). As first observed by \cite{wozniak_microlensing_1997}, in the $f_s\ll1$ and $u_0\ll1$ regime where the baseline flux is dominated by the blend flux, there is strong degeneracy between the three parameters for 1L1S events. While the binary perturbations break some of that degeneracy, strong covariances remain.

\subsection{Binary-Planetary Degeneracy}
\label{sec:binary-deg}

We also provide a fascinating 5-fold-degenerate example that is similar to the degeneracy reported in \cite{choi_new_2012} where a light curve that is blunt and flat near the peak can be explained by either a binary case or a planetary case. Here, we simulate a close-topology, planetary mass ratio ($q=10^{-1.7}$) event where the source trajectory passes through the negative perturbation region towards the back end of the arrowhead-shaped central caustic as in the case of the ``planetary A/B'' caustic in Figure \ref{fig:posterior-binary}. \cite{choi_new_2012} noted that a similar perturbation can occur for the binary case when the source trajectory passes through the negative perturbation region between two adjacent cusps of the astroid-shaped central caustic, as in case of the ``binary A/B/C'' caustics in Figure \ref{fig:posterior-binary}; also see Figure 1 in their paper.

As shown in Figure \ref{fig:posterior-binary}, all five degenerate solutions cause magnification patterns that are hardly distinguishable. The two planetary solutions exhibit a close-wide degeneracy. For the three binary solutions, the ``binary B/C'' solutions suggest two possible trajectories ($\sim90/270 \deg$) for the same lens system configuration whereas ``binary A'' solution exhibits a smaller mass ratio and a wider binary separation than ``binary B/C''. We note that this additional degeneracy in the mass ratio for the binary case was not reported in \cite{choi_new_2012}. It is not clear if this is a discrete or continuous degeneracy, nor if it is an ``accidental degeneracy'' that arises because of the relatively weak perturbation, or is due to some underlying symmetry in the binary lens equation (e.g. \citealt{dominik_binary_1999}).

On the other hand, wide solutions for the binary case are largely absent from the NDE posterior, apart from an inkling of density near $\log_{10}(s)\sim0.69$ which points to the expected close-wide degeneracy for the binary solution. We note that the reason those degenerate solutions are excluded is that, because of the offset between the COM and the central caustic (Equation \ref{eq:offset}), wide-binary solutions would require $t_0<0$, which has a prior probability of zero.

\subsection{Evaluating Performance}
\label{sec:scatter}

\begin{figure*}
 \begin{center}
    \includegraphics[width=\textwidth]{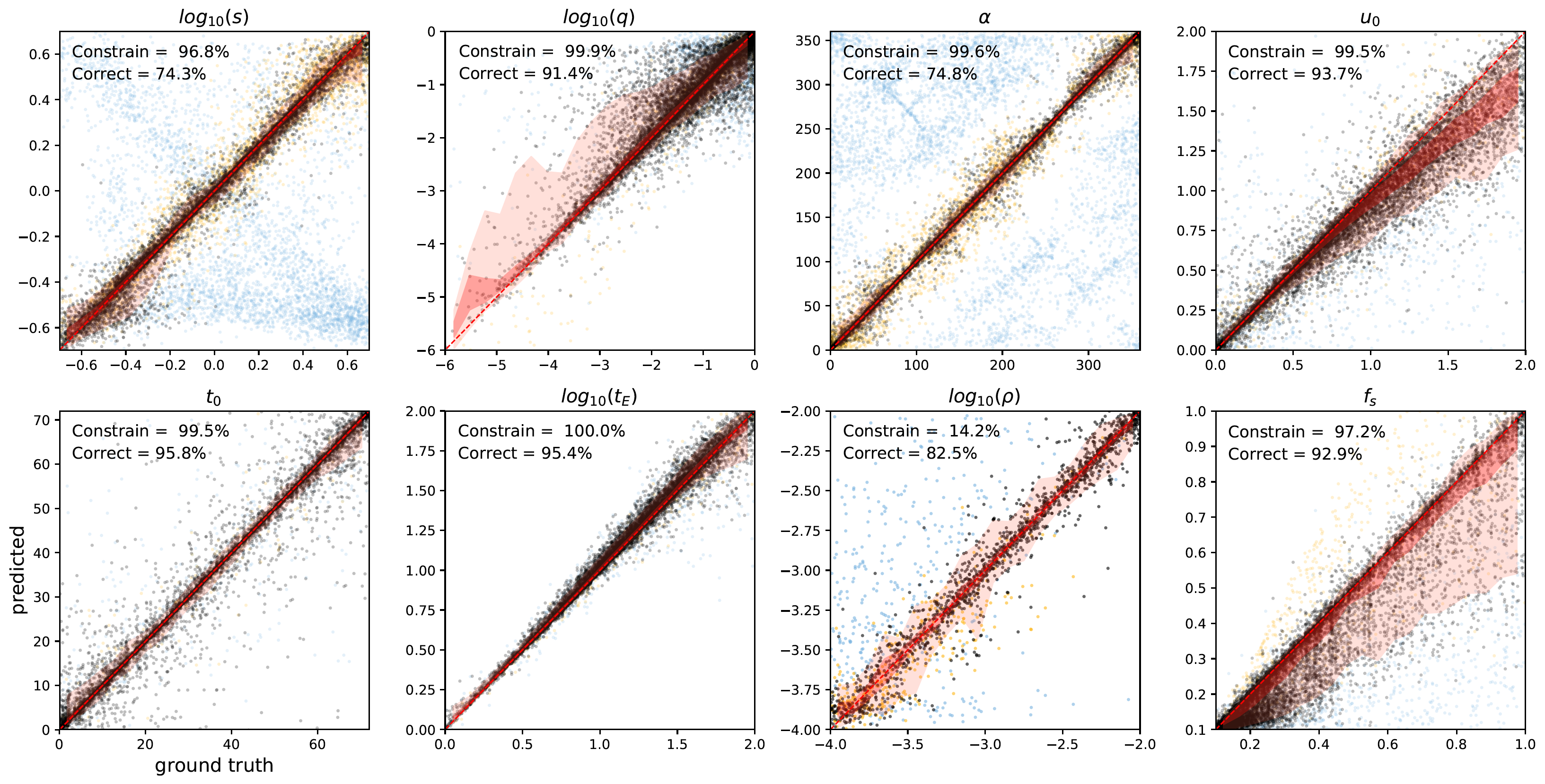}
    \caption{Predicted vs. ground truth 2L1S parameters for 14,551 test-set 2L1S events. $t_E$ and $t_0$ are in units of days, $\alpha$ in degrees, $u_0$, $s$, and $\rho$ in units of $\theta_E$. Single-mode NDE posteriors are shown in black dots. For multi-model NDE posteriors, we color-code the solution as follows: those for which the global mode is closest to the ground truth are plotted in black; for cases where a minor mode is closest to the true value, this correct, minor mode is plotted in orange whereas the incorrect global mode is plotted in blue. Red shadows indicate 32--68th percentile ($1\sigma$) and 5--95 percentile ($2\sigma$) regions. Red-dashed lines show the diagonal. In the upper left of each subplot, ``constrain'' refers to the percentage of events whose NDE posterior poses sufficient constraint---the peak posterior probability must be at least twice the prior probability. ``Correct'' refers to the percentage of constrained events whose true parameter lies closest to the global mode. 
    }
    \label{fig:scatter}
    \end{center}
\end{figure*}

\begin{figure}
    \includegraphics[width=\linewidth]{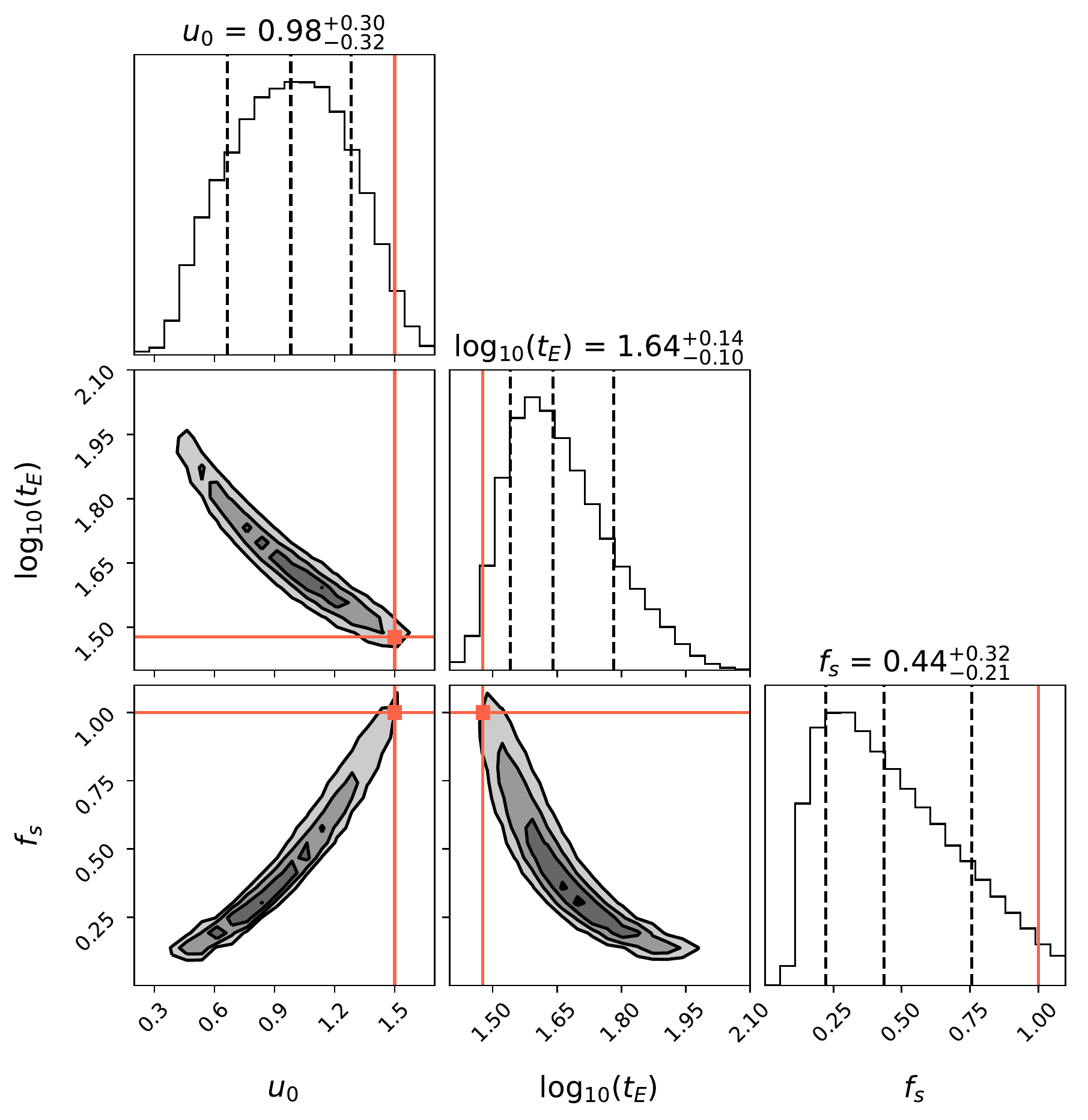}
    \caption{Corner plot for the marginal NDE posterior of an 1L1S event showing strong degeneracy among the three 1L1S parameters: $u_0$ in units of $\theta_E$, $t_E$ in units of days, and $f_s$. Filled contours show 1/2/3/4$\sigma$ regions. Small $u_0$ and $f_s$ are strongly favored because of the effective priors (Section \ref{sec:prior}) and a marginally informative likelihood.}
    \label{fig:1l1s}
\end{figure}

We present a systematic evaluation of all 14,551 test set events in the form of predicted vs.\ truth scatter plots (Figure \ref{fig:scatter}). Each test event light-curve is realized in the same fashion as training time. As the NDE returns potentially multi-modal posteriors of arbitrary shape, we compute the mode(s) for the marginal 1D distributions of the posterior and consider the mode closest to the ground truth as the ``predicted'' value. The mode(s) is computed by first fitting each with a 1D histogram of 100 bins and then searching for local maxima defined as any bin count higher than that of the 20 adjacent bins. This limits the number of modes to 5. Considering the purpose of the NDE posterior is to allow ultra-fast convergence of a downstream sampling-based algorithm like MCMC to determine the exact posterior, as long as the correct solution has substantial density in the NDE posterior, it should not raise alarm if an alternative mode is mistakenly favored. Any degeneracies can be easily resolved downstream. Therefore, it is sensible to allow the correct mode to be used as the predicted value, even if another degenerate mode is incorrectly preferred.

As shown in the upper-left corner of each subplot in Figure \ref{fig:scatter}, all parameters are constrained at a rate of close to $100\%$ except for the finite source effect for which only $14.2\%$ is constrained, as the source trajectory is required to either cross or pass close to a caustic for $\rho$ to be determined.\footnote{Formally, effects on the light curve due to the finite size of the source are only significant if the gradient of the magnification across the source has a significant second derivative. In practice, this condition is only satisfied if the source passes within a few angular source radii of a caustic.}
We consider a parameter to be constrained if the probability density of the 1D marginal distribution is more than twice the prior probability density at the global mode.

The second row in each upper-left corner shows the frequency for which the correct mode is preferred by NDE, that is, the ground truth is the closest to the major mode compared to the minor modes(s), if any. If the ground truth is closer to a minor mode, the major mode is plotted in blue while the major mode is shown in orange. We see clear degeneracy patterns in $\log_{10}(s)$ and $\alpha$. For $\log_{10}(s)$, the ``wide-close'' degeneracy is exhibited by the cluster around the upper-left to lower-right diagonal. For $\alpha$, there is also a cluster of events along the same diagonal, indicating a degeneracy between $\alpha$ and $-\alpha$. Such a degeneracy may happen for nearly symmetrical central caustics along the direction perpendicular to the lens axis.

The 1-$\sigma$ and 2-$\sigma$ ranges of prediction, shown in red shadows, are clustered closely around the diagonal for most parameters. We emphasize that the loose 1L1S-fitting cutoff ($\chi^2_{1L1S}/\rm dof\sim1$) means many of the test-set light-curves are only weakly perturbed by the binary nature of the lens, and should explain a number of cases in which the mass-ratio is poorly constrained. Interestingly, we find that there is a tendency to overestimate the mass ratio in these cases. In addition, we notice that $u_0$ and $f_s$ are underestimated for a large number of cases while $t_E$ is correspondingly overestimated, though hardly visible in Figure \ref{fig:scatter}. This bias could be explained by the combined effect of a known degeneracy for 1L1S events and a distribution mismatch.

First, there exists a well-known degeneracy between $u_0$, $f_s$, and $t_E$ for single-lens events which in our case, applies to events that are only weakly perturbed by the binary nature of the lens. As demonstrated by \cite{wozniak_microlensing_1997}, this degeneracy is most severe for low magnification events ($u_0\gg1$), which is precisely where the biases occur as seen in Figure \ref{fig:scatter}. Indeed, restricted to test events with $u_0<0.15$, the bias in $f_s$ and $t_E$ is largely removed.
Figure \ref{fig:1l1s} shows the NDE posterior for an example $u=1.5$ 1L1S event which demonstrates the strong degeneracy among $u_0$, $f_s$, and $t_E$.

In the presence of strong degeneracies as such, the effective likelihood implicitly provided by the featurizer is only marginally informative. In other words, the featurizer cannot distinguish among solutions within the continuous degeneracy, and only prescribes a region in parameter space where the observation is about equally likely. Therefore, the posterior 
is essentially dominated by the prior, which strongly favors small $u_0$ and $f_s$, as seen in Figure \ref{fig:1l1s}. Had the parameters for the weakly perturbed events in the test-set been drawn from the same effective prior as the full training set, there would be little bias (under/over-estimation) at all in Figure \ref{fig:scatter}. However, quite the contrary, the distribution of the weakly-perturbed is weighted towards the exact opposite direction of effective prior, e.g., towards large $u_0$ and small $\log_{10}(q)$---those more likely to be excluded from the $\chi^2_{1L1S}/\rm dof>1$ cutoff. Because of this distribution mismatch, large $u_0$ and small $\log_{10}(q)$ occur much more often than expected by prior belief, thus resulting in the under/overestimation bias. And because of the strong covariances among $u_0$, $f_s$, and $t_E$, the under-estimation of $u_0$ translates into an under-estimation of $f_s$ and an over-estimation for $t_E$ (Figure \ref{fig:1l1s}), which explains the biases seen in Figure \ref{fig:scatter}.

\begin{figure}
    \includegraphics[width=\linewidth]{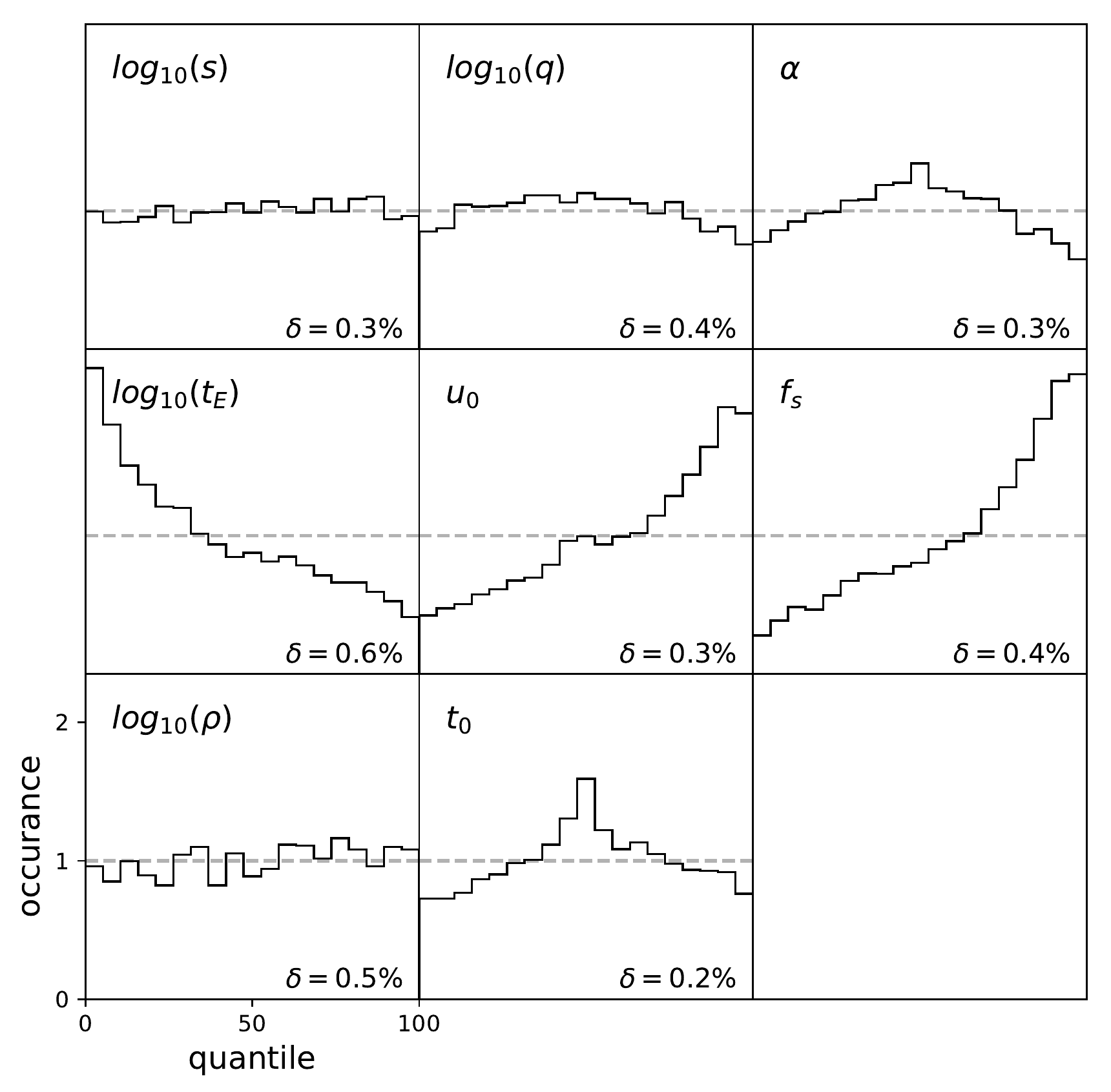}
    \caption{Calibration plot showing the test-set distributions of the ground truth quantile for the 1D marginal NDE posteriors. Dashed lines indicated the uniform distribution as expected for a perfectly calibrated posterior.}
    \label{fig:calibration}
\end{figure}

\subsection{Calibration Properties}

A perfectly calibrated posterior knows how often it is right or wrong. In other words, the quantile of the ground truth parameter under the NDE posterior should be expected to be distributed uniformly.
Figure \ref{fig:calibration} shows the quantile distribution for the 1D marginal NDE posterior distributions for the same 14,551 test set inferences. The quantile distribution for $\log_{10}(q)$, $\log_{10}(\alpha)$, and $t_0$ is concave-up, indicating that the NDE uncertainty is overestimated and the true value lies closer to the center of the posterior more often than expected. This suggests that the NDE finds it hard to contract the posterior in those dimensions, possibly due to numerical optimization difficulties or insufficient neural network expressibility. On the other hand, distributions for the three parameters in the second row---$\log_{10}(t_E)$, $u_0$, and $f_s$---demonstrate the systematic under/over-estimation as seen in Figure \ref{fig:scatter}, where $\log_{10}(t_E)$ is systematically overestimated and $u_0$ and $f_s$ are underestimated. The quantile distributions for $\log(s)$ and $\log_{10}(\rho)$ are consistent with uniform distributions and are thus well-calibrated.

\section{Discussion and Conclusions}
\label{sec:discussion}
We have demonstrated that amortized neural posterior estimation, a likelihood-free inference method which uses a conditional NDE to learn a surrogate posterior, $\hat{p}(\pmb{\theta}|\vect{x})$, greatly accelerates binary microlensing inference---an approximate posterior could be produced in seconds without the need for an expert in the loop. Our new approach is capable of capturing a variety of degeneracies. For future work, it is straightforward to extend to higher-level effects such as parallax and binary motion by introducing additional parameters. Application to more complex systems such as 3L1S may be fruitful, where the physical forward model is orders of magnitude slower. In addition, the photometric noise model adapted in our study is somewhat simplistic, and future work can explore how to adapt models trained with ideal noise properties to fully realistic data with the help of image-based simulation pipelines such as ones used in \cite{penny_predictions_2019}. We discuss two additional aspects of our work below.

\subsection{A hybrid NDE-MCMC framework}
\label{sec:mcmc}
The NDE posterior is easily validated and/or refined by a downstream MCMC sampler. While the NDE posterior is precise enough to allow for fast convergence of downstream MCMC typically within hundreds of steps, we do notice that the precision of the exact MCMC posterior could be more than order-unity higher in many cases. The precision of the NDE posterior is determined by two kinds of uncertainty: data uncertainty and the model uncertainty of the inference algorithm, the latter of which is negligible for MCMC. As neural networks in practice are not infinitely expressive, in the limit of the highest-quality data, the NDE model uncertainty is expected to dominate over data uncertainty. This is the case for \textit{Roman} data. Applied to much noisier and more sparsely sampled ground-based data, we expect that data uncertainties will dominate over model uncertainties, thus allowing the NDE posterior to converge towards the exact posterior.

\subsection{Choice of Coordinate System}
For all events in this work, we have adopted the center-of-mass (COM) coordinate system, which is the default in \texttt{MulensModel} but not the most efficient reference frame in the sense that more than 70\% of the 1 million simulations turn out to be consistent with a 1L1S model. For example, most 2L1S configurations with large $u_0$ do not pass close to either the central caustics or the planetary caustics. For parts of the parameter space, alternative reference coordinates may be more descriptive or useful. For example, the caustic-center frame is preferred for binary and/or wide-separation events for which there is an offset of the caustic-center from the COM. Doing so recovers the missing wide/binary solution in Section \ref{sec:binary-deg} without the need to expand the prior to include negative $t_0$. Additionally, planetary-caustic passing events are also rare; for source trajectories far from the central caustics, most do not pass close to the planetary caustic and as a result, the magnification is frequently indistinguishable from 1L1S. For future work, a hybrid and self-consistent coordinate system could be used.

\software{%
    \texttt{numpy} \citep{vdw11},
    \texttt{scipy} \citep{scipy},
    \texttt{matplotlib} \citep{2007CSE.....9...90H}, 
    \texttt{pytorch} \citep{paszke2017automatic},
    \texttt{MulensModel} \citep{poleski_mulensmodel_2018},
    \texttt{Jupyter} \citep{Kluyver:2016aa},
    \texttt{corner} \citep{foreman-mackey_cornerpy_2016}.
}

\section*{Acknowledgements}

K.Z.\ and J.S.B.\ are supported by a Gordon and Betty Moore Foundation Data-Driven Discovery grant. J.S.B.\ is partially sponsored by a faculty research award from Two Sigma. K.Z.\ thanks the LSSTC Data Science Fellowship Program, which is funded by LSSTC, NSF Cybertraining Grant 1829740, the Brinson Foundation, and the Moore Foundation; his participation in the program has benefited this work. K.Z.\ thanks Shude Mao and Tsinghua University for their hospitality during the COVID-19 pandemic. B.S.G.\ is supported by NASA grant NNG16PJ32C and the Thomas Jefferson Chair for Discovery and Space Exploration. This work is supported by the AWS Cloud Credits for Research program. We thank Yang Gao and Yu Sun for helpful discussions, and Weicheng Zang for pointing out issues with the initial figure presentation.

\end{document}